\documentclass[twocolumn]{aastex63}

\usepackage{amsmath}
\usepackage{graphicx}
\usepackage{natbib}
\usepackage[scaled]{helvet}
\usepackage{epsfig}
\usepackage{url}
\usepackage{textcomp}
\usepackage{mathpazo}
\usepackage{xspace}
\usepackage{apjfonts}
\usepackage{rotating}
\usepackage{needspace}
\usepackage{balance}

\usepackage{color}
\usepackage[normalem]{ulem}

\usepackage{units} %nicefrac

\bibpunct{(}{)}{;}{a}{}{,}
\newcommand\mysim{\mathord{\sim}}

\newcommand\dg{\ensuremath{^\circ}}
\def\lsim{\mathrel{\rlap{\lower4pt\hbox{\hskip1pt$\sim$}}
    \raise1pt\hbox{$<$}}}
\def\gsim{\mathrel{\rlap{\lower4pt\hbox{\hskip1pt$\sim$}}
    \raise1pt\hbox{$>$}}}

\newcommand{\rffigl}[1]{Figure~\ref{fig:#1}}
\newcommand{\rffigs}[2]{Figures~\ref{fig:#1} and \ref{fig:#2}}

\newcommand{\rfsecl}[1]{\mbox{Section \ref{sec:#1}}}

\newcommand{\rftabl}[1]{Table~\ref{tab:#1}}

\newcommand{\rfeql}[1]{Equation~\ref{eq:#1}}

\newcommand\teff{\ifmmode{T_{\rm eff}}\else $T_{\rm eff}$\fi}
\newcommand\logg{\ifmmode{\log{g}}\else $\log{g}$\fi}
\newcommand{\feh}{\mbox{$\rm{[Fe/H]}$}}
\newcommand{\mh}{\mbox{$\rm{[m/H]}$}}
\newcommand\vsini{\ifmmode{v\sin{i_\star}}\else $v\sin{i_\star}$\fi}

\newcommand\sini{\ifmmode{\sin{i_\star}}\else $\sin{i_\star}$\fi}
\newcommand\msun{\ifmmode{M_{\odot}}\else $M_{\odot}$\fi}
\newcommand\rsun{\ifmmode{R_{\odot}}\else $R_{\odot}$\fi}
\newcommand\lsun{\ifmmode{L_{\odot}}\else $L_{\odot}$\fi}
\newcommand\mstar{\ifmmode{M_{\star}}\else $M_{\star}$\fi}
\newcommand\rstar{\ifmmode{R_{\star}}\else $R_{\star}$\fi}
\newcommand\mjup{\ifmmode{M_{\rm Jup}}\else $M_{\rm Jup}$\fi}
\newcommand\rjup{\ifmmode{R_{\rm Jup}}\else $R_{\rm Jup}$\fi}
\newcommand\mearth{\ifmmode{M_\oplus}\else $M_\oplus$\fi}
\newcommand\rearth{\ifmmode{R_\oplus}\else $R_\oplus$\fi}
\newcommand\mpl{\ifmmode{M_{\rm P}}\else $M_{\rm P}$\fi}
\newcommand\mb{\ifmmode{M_{\rm b}}\else $M_{\rm b}$\fi}
\newcommand\rp{\ifmmode{R_{\rm P}}\else $R_{\rm P}$\fi}
\newcommand\pchi{\ifmmode{P(\chi^2)}\else $P(\chi^2)$\fi}
\newcommand\aen{\ifmmode{\sigma_\pi}\else $\sigma_\pi$\fi}
\newcommand\kms{\ifmmode{\rm km\thinspace s^{-1}}\else km\thinspace s$^{-1}$\fi}
\newcommand\ms{\ifmmode{\rm m\thinspace s^{-1}}\else m\thinspace s$^{-1}$\fi}
\newcommand\prot{\ifmmode{P_{\rm rot}}\else $P_{\rm rot}$\fi}
\newcommand\kep{{\em Kepler}}
\newcommand{\ktwo}{{\em K2}}
\newcommand\tess{{\em TESS}}
\newcommand\corot{{\em CoRoT}}
\newcommand\elec{\mathrm{e^-}}

\newcommand{\bjdtdb}{\ensuremath{\rm {BJD_{TDB}}}}

\newcommand{\fave}{\langle F \rangle}
\newcommand{\favee}{\langle F_\oplus \rangle}

\newcommand{\thisstar}{KIC4918810\xspace}
\newcommand{\thisstarb}{KIC4918810\,b\xspace}

\DeclareMathOperator\erf{erf}
\defcitealias{fortney:2007}{F07}

\newcommand{\citeta}{\citetalias}

\newcommand{\cfa}{Center for Astrophysics | Harvard \& Smithsonian, 60 Garden St, Cambridge, MA 02138, USA}

\newcommand{\MIT}{Department of Physics and Kavli Institute for Astrophysics and Space Research, Massachusetts Institute of Technology, Cambridge, MA 02139, USA}

\newcommand{\berkeley}{Astronomy Department, University of California Berkeley, Berkeley, CA 94720-3411, USA}

\newcommand{\bishops}{Department of Physics and Astronomy, Bishop's University, 2600 College St., Sherbrooke, QC J1M 1Z7, Canada}
\newcommand{\caltech}{Department of Astronomy, California Institute of Technology, Pasadena, CA 91125, USA}
\newcommand{\wisconsin}{Department of Astronomy, University of Wisconsin-Madison, Madison, WI 53706, USA}

\newcommand{\amateur}{\altaffiliation{Amateur Astronomer}}

\begin{document}

\title{A Long-period Substellar Object Exhibiting a Single Transit in \kep}

\correspondingauthor{Samuel N. Quinn}
\email{squinn@cfa.harvard.edu}

\author[0000-0002-8964-8377]{Samuel N. Quinn}
\affiliation{\cfa}

\author[0000-0003-3182-5569]{Saul Rappaport}
\affiliation{\MIT}

\author[0000-0001-7246-5438]{Andrew Vanderburg}
\affiliation{\wisconsin}

\author[0000-0003-3773-5142]{Jason D. Eastman}
\affiliation{\cfa}

\author[0000-0002-6916-8130]{Lorne A. Nelson}
\affiliation{\bishops}

\author[0000-0003-3988-3245]{Thomas L. Jacobs}
\amateur
\affiliation{12812 SE 69th Place Bellevue, WA 98006, USA}

\author[0000-0002-8527-2114]{Daryll M. LaCourse}
\amateur
\affiliation{7507 52nd Place NE Marysville, WA 98270, USA}

\author{Allan R. Schmitt}
\amateur
\affiliation{616 W. 53rd. St., Apt. 101, Minneapolis, MN 55419, USA}

\author{Perry Berlind} 
\affiliation{\cfa}

\author[0000-0002-2830-5661]{Michael L. Calkins} 
\affiliation{\cfa}

\author[0000-0002-9789-5474]{Gilbert A. Esquerdo} 
\affiliation{\cfa}

\author[0000-0001-8638-0320]{Andrew W. Howard}
\affiliation{\caltech}

\author[0000-0002-0531-1073]{Howard Isaacson}
\affiliation{\berkeley}

\author[0000-0001-9911-7388]{David W. Latham}
\affiliation{\cfa}

% \author{others?}
% \noaffiliation{}

\shorttitle{A \kep\ single transit}
\shortauthors{Quinn et al.}

\begin{abstract}

We report the detection of a single transit-like signal in the \kep\ data of the slightly evolved F star \thisstar. The transit duration is $\mysim45$\ hours, and while the orbital period ($P\mysim10$\ years) is not well constrained, it is one of the longest among companions known to transit. We calculate the size of the transiting object to be $\rp = 0.910$\,\rjup. Objects of this size vary by orders of magnitude in their densities, encompassing masses between that of Saturn ($0.3$\,\mjup) and stars above the hydrogen-burning limit ($\mysim80$\,\mjup).  Radial-velocity observations reveal that the companion is unlikely to be a star. The mass posterior is bimodal, indicating a mass of either $\mysim0.24$\,\mjup\ or $\mysim26$\,\mjup. Continued spectroscopic monitoring should either constrain the mass to be planetary or detect the orbital motion, the latter of which would yield a benchmark long-period brown dwarf with a measured mass, radius, and age. \\

\end{abstract}

\section{Introduction} 

To date, the most successful techniques for finding planets suffer from detection biases against long-period companions. The radial velocity (RV) semi-amplitude scales with $P^{-\nicefrac{1}{3}}$, which limits the detection of long-period planets to only the most massive objects. Among planets with minimum masses less than that of Saturn, for example, only three have been discovered via RVs with an orbital period longer than $5$\ years \citep[HD\,10180\,h, GJ\,15\,A\,c, GJ\,433\,c;][]{lovis:2011,pinamonti:2018,feng:2020}. Moreover, to fully characterize a planet, one needs additional information, such as size constraints that come from a transit, the probability of which scales inversely with semi-major axis, leading to a dependence of $P^{-\nicefrac{2}{3}}$. The duty cycle of transit observations is also a limiting factor, as all of the information about the presence of a long-period companion comes from the transit itself (whereas RV signal is distributed throughout the orbit). Ground-based transit surveys, limited by night-time and seasonal observability, therefore have trouble detecting companions with orbital periods longer than about a week. The longest-period planet from a ground-based transit survey is LHS \,1140\,b \citep[$P=24.7$\ days;][]{Dittmann:2017}. While much can be learned by studying the shortest-period transiting planets, they may not form or evolve in the same ways as their long-period counterparts. Other techniques (e.g., microlensing and direct imaging), have detected planets more widely separated from their host stars, but transits are still required to independently measure planetary radii.

Fortunately, space missions like \kep\ and \ktwo\ \citep{borucki:2010,Howell:2014}, \corot\ \citep{auvergne:2009}, and \tess\ \citep{Ricker:2015} gather observations of large numbers of stars with a much higher duty cycle and long time spans, so they are capable of detecting long-period transiting planets. \kep\ in particular has observed a handful of planets and planet candidates beyond 1 AU \citep[e.g.,][]{wang:2015}; the longest-period transiting planet with a mass and a well determined period is the circumbinary planet Kepler-1647b \citep[$P=1100$\,days; $\mpl=1.5\,\mjup$;][]{kostov:2016}. For periods longer than the $4$-year \kep\ observing time span, however, only a single transit can be observed. 

Some of the challenges posed by single-transit systems were explored in \citet{yee:2008}, who predicted at least a handful of single-transit giant planets from \kep\ and described a method for efficient RV follow-up. The biggest difference for single-transit systems compared to other transiting companions is of course the unknown orbital period. When the period is known, the transit duration can be used to constrain the combination of mean stellar density, impact parameter, and orbital eccentricity. For single transit systems, the eccentricity is degenerate with the orbital period and largely unconstrained. When the period is known, the phase of the radial velocity measurements directly constrain the companion mass and eccentricity. For single transits, the phase of the observations is not known and RVs provide a weaker constraint. Nevertheless, single transits provide an opportunity: for all transit surveys to date the uninterrupted time span of photometric observations is shorter than the orbital periods of some particularly interesting classes of object. For \tess, single observing sectors are only $\mysim27$\ days long. For \ktwo, campaigns were no more than $\mysim 80$\ days. In both surveys, planets transiting in the habitable zones of many stars would only exhibit a single transit \citep[see, e.g.,][]{vanderburg:2018}. In its two-year prime mission, \tess\ was predicted to observe $\mysim 1000$\ single transits, including $\mysim 80$\ in the habitable zones of their stars \citep{villanueva:2019}, the search for which is ongoing. During the four-year prime \kep\ mission, planets in the habitable zones of Sun-like stars might show multiple transits, but objects orbiting on Solar System scales of several AU would be observed to transit no more than once. Gas giant planets with orbital periods close to a decade---Jupiter analogs---have been detected by radial velocity surveys \citep[e.g.,][]{wittenmyer:2020,rosenthal:2021}, and their ensemble orbital properties place constraints on typical planetary system architectures, but their physical properties remain unconstrained because without transits (or potentially direct imaging), we cannot know their sizes, their atmospheric compositions, or their physical evolution. Even more widely separated planets and brown dwarfs have been directly imaged, but some of their physical properties can only be constrained by models, and only if their ages and temperatures can be determined \citep[see, e.g.,][]{jones:2016}. Transits of long-period companions would allow us to determine their sizes and compositions, and to more directly compare the properties of well characterized short-period planets to those of the long-period RV and directly imaged planets. If the host star is close enough, transiting planets with long periods may even be amenable to characterization through direct imaging or astrometry.

In this paper, we present the \kep\ discovery and initial characterization of a transiting companion to the star \thisstar, which is among the brightest hosts of such objects. We describe the \kep\ photometry and follow-up spectroscopy in \rfsecl{data}, the global fit in \rfsecl{exofast}, and we discuss future prospects for the system---and systems like it that should be discovered by \tess---in \rfsecl{discussion}. We note that while we were finishing our follow-up observations, \citet{kawahara:2019} published an independent discovery of this system as part of a comprehensive single transit search in \kep\ data and an examination of the characteristics of the \textit{population} of single transits, rather than characterization of individual systems.

\section{Observations}
\label{sec:data}

\subsection{\kep\ Photometry}

\begin{table}
\scriptsize
\centering
\caption{Literature Properties for \thisstar}
\begin{tabular*}{\columnwidth}{l @{\extracolsep{\fill}} lcr}
\hline
\hline
            & \multicolumn{3}{c}{TIC 122301267} \\
Other       & \multicolumn{3}{c}{2MASS J19210170+4002148} \\
identifiers	& \multicolumn{3}{c}{Gaia DR2 2101102266113728384} \\
            & \multicolumn{3}{c}{APASS 51739229} \\
\hline
\hline
Parameter & Description & Value & Source\\
\hline 
$\alpha_{J2000}$	&R.A.  & 19:21:01.708 & 1	\\
$\delta_{J2000}$	&decl. & +40:02:14.83 & 1	\\
$\mu_{\alpha}$	& PM in R.A. (mas yr$^{-1}$)   & -1.400 $\pm$ 0.023	& 1 \\
$\mu_{\delta}$	& PM in decl. (mas yr$^{-1}$)  & -5.319 $\pm$ 0.025 & 1 \\
$\pi$ & Parallax (mas)  & 0.924 $\pm$ 0.015 & 1 \\
\\[-2ex]
$RV$ & Systemic RV (\kms)   & -3.48 $\pm$ 0.13 & 2 \\
\\[-2ex]
$K_p$   & \kep\ $K_P$ mag & 13.359 $\pm$ 0.022 & 3 \\
\\[-2ex]
$B$     & APASS Johnson $B$ mag	& 14.139 $\pm$ 0.035 & 4	\\
$V$		& APASS Johnson $V$ mag	& 13.403 $\pm$ 0.015 & 4	\\
$g'$	& APASS Sloan $g'$ mag	& 13.722 $\pm$ 0.016 & 4	\\
$r'$	& APASS Sloan $r'$ mag	& 13.326 $\pm$ 0.014 & 4	\\
$i'$	& APASS Sloan $i'$ mag	& 13.225 $\pm$ 0.090 & 4	\\
\\[-2ex]
$G$      & {\it Gaia} $G$ mag   & 13.3272 $\pm$ 0.0002 & 1 \\
$G_{BP}$ & {\it Gaia} $BP$ mag  & 13.6187 $\pm$ 0.0010 & 1 \\
$G_{RP}$ & {\it Gaia} $RP$ mag  & 12.8853 $\pm$ 0.0009 & 1 \\
\\[-2ex]
$T$	     & \tess\ $T$ mag	    & 12.9415 $\pm$ 0.0076 & 5 \\
\\[-2ex]
$J$		 & 2MASS $J$ mag        & 12.367 $\pm$ 0.02    & 6 \\
$H$		 & 2MASS $H$ mag        & 12.130 $\pm$ 0.02    & 6 \\
$K_S$    & 2MASS $K_S$ mag      & 12.077 $\pm$ 0.02    & 6 \\
\\[-2ex]
\textit{W1}		& WISE \textit{W1} mag & 12.077 $\pm$ 0.023 & 7	\\
\textit{W2}		& WISE \textit{W2} mag & 12.112 $\pm$ 0.022 & 7 \\
\textit{W3}		& WISE \textit{W3} mag & 12.586 $\pm$ 0.460	& 7	\\
\hline
\\[-6ex]
\end{tabular*}
\begin{flushleft} 
\footnotesize{\vspace{6pt}
% {\bf Note.\\}
% $^{\rm a}$ The uncertainties of the photometry have a systematic error floor applied. However, the global fit requires a significant scaling of the uncertainties quoted here to be consistent with our model, suggesting they are still significantly underestimated for one or more of the broadband magnitudes.\\[0.25ex]
    {\bf References.} (1) \citet{Gaia:2018}; (2) this work; (3) \citet{brown:2011}; (4) \citet{Henden:2016}; (5) \citet{stassun:2018b}; (6) \citet{Cutri:2003}; (7) \citet{Cutri:2014}.
}
\end{flushleft}
\label{tab:lit}
\end{table}

\begin{figure*}[!htb]
    \centering
    \includegraphics[width=\linewidth]{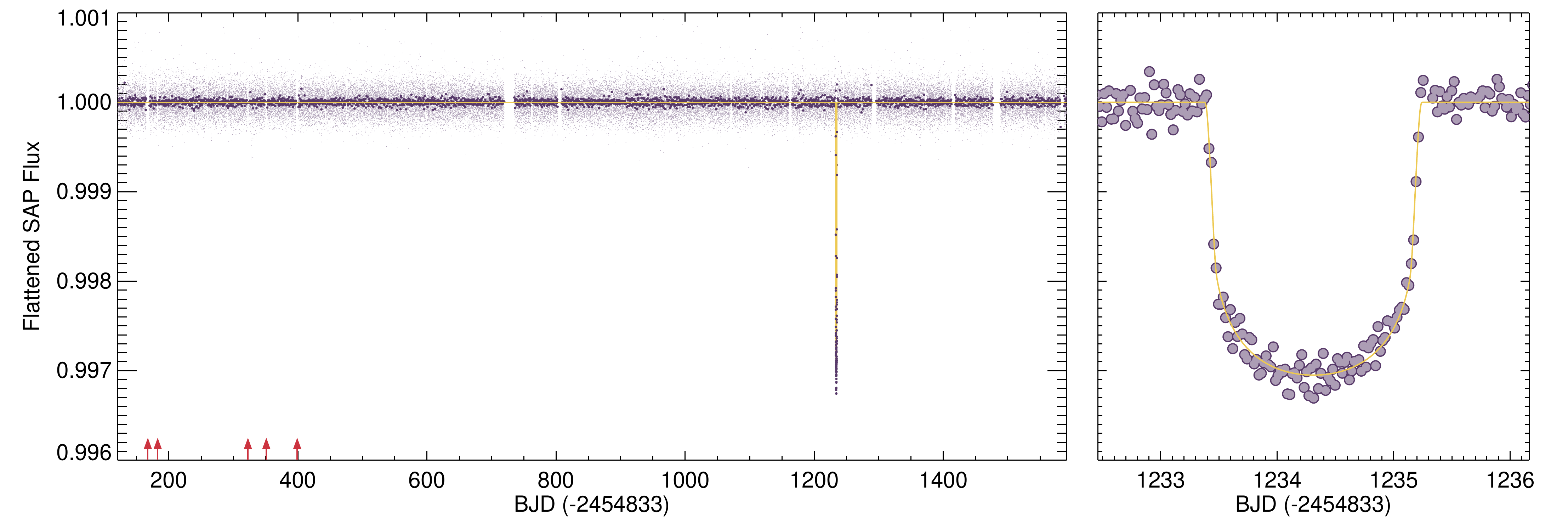}
    \caption{The detrended, normalized, \kep\ light curve of \thisstar. The full \kep\ data set is presented in the left panel, with light purple denoting individual long cadence data points and dark purple circles denoting the same data in 12-hour bins as well as the in-transit data for clarity. The best-fit transit model is plotted in yellow. We also note $5$ gaps in the first half of the \kep\ data set (red arrows) that are longer than the transit duration, and in which a second transit could have possibly fallen. These gaps would correspond to periods of $835$, $883$, $912$, $1052$, and $1067$\ days. On the right, we show a closer view of the transit showing the long cadence data and the best-fit model.}
    \label{fig:lc}
\end{figure*}

\thisstar\ was observed for all quarters (Q0 to Q17) of the prime \kep\ mission (BJD 2454953.5 to 2456424.0), a total of 1470.5 days. The broadband photometric measurements and position of the star are given in \rftabl{lit}. Like most \kep\ targets, it was observed in long cadence mode with a cadence of nearly $30$\ minutes. The characteristics of the detector are described in \citet{koch:2010} and \citet{vancleve:2016}, the data reduction pipeline is described in \citet{jenkins:2010a} and \citet{jenkins:2017}, and the long-cadence photometric performance is described in \citet{jenkins:2010b}. 

We carried out a visual search for single transits in the \kep\ data using the LcTools software\footnote{\url{https://sites.google.com/a/lctools.net/lctools/}} \citep{schmitt:2019} to examine the Simple Aperture Photometry (SAP) Pre-search Data Conditioning (PDC) light curves. PDC light curves use cotrending basis vectors to remove instrumental effects from the light curves and generally result in cleaner data for visual inspection. This search led to the identification of a transit-like event at ${\rm BJD}-2454833 = 1234.3$\ with a duration of nearly $2$\ days. The long duration of the event leads to systematic effects near transit in the PDC reduction, so further analysis of the system used the SAP flux, which we detrended using basis splines while masking out the transit. The detrended and normalized light curve is shown in \rffigl{lc}. From this figure, it is clear by eye that there are no other transits in the data, although there are five gaps longer than the transit duration early in the mission, into which a second transit may have fallen. These five gaps would correspond to orbital periods of roughly $835$, $883$, $912$, $1052$, or $1067$\ days, all of which are possible if allowing arbitrary orbital eccentricities. However, these five discrete periods are many orders of magnitude less likely than a longer period, which does not require finely tuned orbital properties and transit timing.

\subsection{HIRES Spectroscopy}

We obtained one spectrum using the HIRES spectrograph on the Keck I 10-m telescope on Mauna Kea, HI, on UT 2017 Apr 10. The 413-second observation was taken without the iodine cell using the C2 decker ($14\arcsec$\ by $0.861\arcsec$), resulting in a resolving power $R\mysim60,000$.
A spectral analysis using the SpecMatch Routine \citep{petigura:2015} yielded spectroscopic stellar parameters $\teff = 6318 \pm 100$\,K, $\logg = 4.22 \pm 0.10$, $\feh = +0.25 \pm 0.06$, and $\vsini = 11$\ \kms. The spectrum was searched for secondary spectral lines \citep{kolbl:2015} resulting in no detections of companion stars down to $\pm 10$\ \kms\ of the systemic velocity of the primary with $>2$\% of its brightness. The systemic radial velocity was measured at $-3.5 \pm 0.2$\ \kms\ \citep{chubak:2012}.

\subsection{TRES Spectroscopy} \label{sec:tres}

We obtained $13$\ high resolution spectra with the Tillinghast Reflector Echelle Spectrograph \citep[TRES; ][]{furesz:2008} between UT 2017 May 05 and 2019 May 31. TRES is mounted on the 1.5-m Tillinghast Reflector at the Fred L. Whipple Observatory on Mount Hopkins, AZ. It has a resolving power of $R\mysim44,000$, and a wavelength coverage of $3850$--$9100$\ {\AA}. At $V=13.4$, \thisstar\ is faint for the 1.5-m, but TRES has the advantage of low read noise ($2.7\,\elec$), which allows good radial velocity performance even at low SNR. However, these observations are still susceptible to sky contamination. Three of the observations were obtained in sub-optimal observing conditions with bright moon, and they show evidence for lunar contamination, such as line width and line profile changes consistent with the presence of blended solar spectrum at the velocity appropriate for lunar contamination, and the expected (spurious) RV shifts that would result from that contamination. We therefore removed them from our data set. The $10$\ remaining spectra have signal-to-noise ratios (SNR) between 17 and 28 per resolution element.

We optimally extract the spectra and run multi-order cross-correlations to derive radial velocities following the process outlined in \citet{Buchhave:2010}, with the exception of the template. Rather than use the strongest observed spectrum as a template, we perform an initial RV analysis to shift and median combine the spectra to produce a template with ${\rm SNR}=75$. We also use this template to perform cosmic ray removal, replacing the affected wavelengths with the median spectrum instead of a linear interpolation. 

TRES is not pressure-controlled, and the instrumental zero point can vary over time by a few tens of meters per second. For observations of \thisstar, for which a slow drift of a few tens of meters per second could mean the difference between a stellar and substellar companion, this is an important detail. We track the zero point of TRES using nightly observations of several standard stars \citep[see, e.g.,][]{quinn:2012b}, and the zero point shifts are typically determined to better than a few meters per second. The standard deviation of the zero-point-corrected standard star RVs also provides an estimate of the per-point instrumental precision, $\sigma_{\rm TRES}$, which we find to be $\mysim12$\,\ms\ at most epochs. While the formal uncertainties of $\mysim 65$\,\ms\ dominate the error budget for \thisstar, we add $\sigma_{\rm TRES}$\ in quadrature nonetheless. The zero-point-corrected RVs with total uncertainties are presented in \rftabl{rv}.

\begin{table}
\small
\centering
\caption{TRES RVs of \thisstar}
\begin{tabular*}{\columnwidth}{l @{\extracolsep{\fill}} rr}
% \begin{tabular*}{0.8\columnwidth}{lrr}
\hline
\hline
\bjdtdb & RV (m s$^{-1}$) & $\sigma_{RV}$ (m s$^{-1}$) \\
\hline
\\[-2.5ex]
2457880.940672    &      -8.9    &   62.4 \\
2457933.725318    &     -19.3    &   54.0 \\
2457994.857476    &     -31.0    &   62.0 \\
2458205.997178    &      -0.6    &   79.8 \\
2458302.825049    &     141.7    &   55.5 \\
2458306.718811    &      46.6    &   72.9 \\
2458331.862689    &       6.1    &   44.4 \\
2458597.874932    &      97.3    &   67.5 \\
2458605.929313    &      48.5    &   54.0 \\
2458634.943225    &      96.4    &   77.2 \\
\\[-2.5ex]
\hline
\\[-5.5ex]
\end{tabular*}
\begin{flushleft} 
\footnotesize{\vspace{6pt}
    {\bf Note.} Relative TRES radial velocities, derived via cross-correlation against the median-combined template and shifted to account for changes in the TRES zero point, as described in \rfsecl{tres}.
    }
\end{flushleft}
\label{tab:rv}
\end{table}

We run the Stellar Parameter Classification software \citep[SPC;][]{Buchhave:2012} on the median TRES spectrum as an independent check on the HIRES classification. We find $\teff = 6230 \pm 50$\,K, $\logg = 4.19 \pm 0.10$, $\mh = 0.231 \pm 0.080$, and $\vsini = 12.0 \pm 0.5$\,\kms. These parameters are consistent with those from HIRES, and derived from a spectrum with similar resolution but greater SNR. For this reason, we choose to use the SPC results in the subsequent analysis of the system parameters. However, because of known degeneracies in the spectroscopic determination of \teff, \logg, and \mh, we perform an iterative fit between SPC and the SED (see \rftabl{lit}) to arrive at a self-consistent, accurate set of parameters. We use \texttt{EXOFASTv2} to fit the SED and Gaia parallax while placing a prior on the metallicity from SPC. This yields a surface gravity of $\logg=4.048 \pm 0.040$, which we then fix in a second SPC analysis, yielding $\teff = 6130 \pm 50$\,K and $\mh=0.143 \pm 0.080$. The final stellar parameters will come from the joint model described in \rfsecl{exofast}, and in which we set a prior on metallicity based on these SPC results.

\subsection{Imaging}
\label{sec:imaging}

High resolution imaging might plausibly be used to detect a long-period stellar companion that is the source of the transit. Unfortunately, at 1 kpc, $0.1\arcsec$\ resolution would only be sensitive to companions at separations greater than $100$\ au, whereas the transit duration for \thisstarb\ corresponds to $a \mysim 5$\,au for a circular orbit. The other reason to obtain high resolution imaging is to ensure the transit depth is not diluted by the light from a close stellar companion. However, any companion bright enough to significantly influence the transit depth should also be apparent as a problem in the global model. A cooler companion would appear as a second component in the SED (manifesting as a poor fit to a single-star model), while a similar-temperature companion would lead to tension between the photometric and astrometric parallax (manifesting as a stellar radius larger than implied by the observed spetroscopic parameters and isochrones). Either case might also induce Gaia DR2 astrometric excess noise \citep{lindegren:2018a}, or the recommended goodness-of-fit statistic, the Renormalized Unit Weight Error \citep[RUWE;][]{lindegren:2018b}. We observe none of these problems in the fit or the data, and conclude that there is no significant flux contribution from unresolved stars.

\section{EXOFAST\lowercase{v}2 Global Fit} 
\label{sec:exofast}

We use the \texttt{EXOFASTv2} global fitting software \citep{Eastman:2017,eastman:2019} to perform a joint fit of the \kep\ photometry within three days of the single transit event, TRES radial velocities, Gaia DR2 parallax, broad-band literature photometry, and the spectroscopic stellar parameters in conjunction with the MESA Isochrones \& Stellar Tracks \citep[MIST;][]{paxton:2011,dotter:2016,choi:2016}. We first iteratively fit the SED and the spectroscopy, as described in \rfsecl{tres}, to break the degeneracy between spectroscopic parameters, and we set Gaussian priors using the resulting SPC metallicity and \teff. We do not set a prior on \logg, instead letting it be constrained by the SED, parallax, and MIST models. We set a Gaussian prior on the parallax from Gaia DR2, and we include a uniform prior on the extinction, $0 \leq A_V \leq 0.36$, requiring it to be less than the mean of the estimates of the total line of sight extinction from \citet{schlafly:2011} and \cite{Schlegel:1998}. We allow all other parameters to be fit without prior constraint, but we note that not all solutions are equally likely or, in some cases, even physically plausible. Such situations can be addressed via rejection sampling, the process of using prior knowledge of the likelihood of a value to reject (with an appropriate probability) solutions that are unlikely. \citet{blunt:2017} presented a framework for using rejection sampling to fit sparsely sampled astrometric orbits, and \citet{vanderburg:2018} showed one way in which rejection sampling can be used to narrow the period posterior of single transits. In Sections \ref{sec:single} through \ref{sec:gaia}, we describe the rejection sampling that we apply to winnow down the \texttt{EXOFASTv2} posterior, including rejection based on transit probabilities, eccentricity distributions, mass-radius relationships, and unresolved astrometric orbits. While we ultimately reject a large fraction of our posterior in this process, we compensate by enforcing conservative convergence criteria in the fit---requiring $>3000$\ independent draws and a Gelman-Rubin statistic $<1.01$ for each parameter. As a result, the chains still contain $>100$\ independent draws (and usually many more) for every parameter. Following discussion in \citet{eastman:2019}, we therefore expect our derived parameters to be accurate to (much) better than $\mysim0.1\sigma$.

In addition to the fully unconstrained fit, we also explore a circular fit, which allows us to more directly compare results with some previous single transit systems \citep[e.g.,][]{giles:2018}. We comment on the eccentric solutions and discuss our results, along with the impact of future observations, in Sections \ref{sec:ew} through \ref{sec:discussion}.

\subsection{Period prior for single transits}
\label{sec:single}

Long-period orbits are less likely to transit in a fixed time window simply because the duty cycle is lower than for short-period orbits. This allows us to apply a prior to the orbital periods following the method described by \citet{vanderburg:2018}. That is, for orbital periods longer than the observing baseline, $\Delta T$, the probability of observing a transit decreases as $1/P$, so we reject draws from the posterior according to \rfeql{single}.
\begin{equation}\label{eq:single}
    {\Pr}_{rej}(P) = 1 - (\Delta T + t_{14})/P
\end{equation}
where $t_{14}$\ is the full transit duration. The effect of this prior is to appropriately down-weight long-period orbits in the posterior.

We also note that with the exception of the five short gaps in the light curve during which a second transit could have occurred (see \rffigl{lc}), periods shorter than 1114.5 days can be excluded. We place no prior constraint on the orbital period in the \texttt{EXOFASTv2} fit, and instead reject these short-period draws afterward. Only $0.07\%$ of the posterior ultimately lands in the gaps in the light curve; the period is likely much longer.

\subsection{Eccentricity prior for long-period planets}
\label{sec:beta}

\citet{Kipping:2013} found that the eccentricity distribution of known exoplanets is well described by a Beta distribution
\begin{equation}
    {\Pr}_\beta(e;a,b) = \frac{1}{B(a,b)}e^{a-1}(1-e)^{b-1}
\end{equation}
\noindent where $B(a,b)$ is the Beta function, with $a = 1.12^{+0.11}_{-0.10}$\ and $b = 3.09^{+0.32}_{-0.29}$\ for long-period planets ($P>382$\ days). Following this work, we draw eccentricities from the distribution
\begin{equation}\label{eq:edist}
\begin{aligned}
    {\Pr}_\beta(e) =&~\frac{1}{B(1.12,3.09)} \times e^{0.12}(1-e)^{2.09} \\
               =&~3.827 \times e^{0.12}(1-e)^{2.09}
\end{aligned}
\end{equation}
\noindent This is equivalent to rejecting draws with the complementary probability
\begin{equation}
    {\Pr}_{rej}(e) = 1 - \big(3.827 \times e^{0.12}(1-e)^{2.09}\big).
\end{equation}
The eccentricity distribution described in \rfeql{edist} peaks at low eccentricity, and is not terribly approximated (modulo normalization) by the simple linear relationship $\Pr(e) \mysim 1 - e$.

We also require that the periastron distance, $a (1 - e)$, be greater than $3\,\rstar$, so we reject draws for which $e > 1 - (\nicefrac{3\,\rstar}{a})$. Given the long orbital period, this only removes the most extreme eccentricities ($e \gsim 0.995$). The orbits of planets with such high eccentricities would be expected to rapidly circularize and shrink through tidal dissipation; these orbits are incompatible with the estimated age of the system of $3$\,Gyr.

\subsection{Mass-radius priors}
\label{sec:mr}

While objects the size of \thisstarb\ may be planets, brown dwarfs, or low-mass stars, not {\it all} objects with masses between that of Saturn and the lower main sequence are consistent with the measured radius ($0.910\,\rjup$\ from the \texttt{EXOFASTv2} fit described above). As mass is added to the envelopes of giant planets, the radius of the planet increases until roughly $5$\,\mjup, with the exact turnover point dependent on core mass, orbital separation, and age of the system \citep[see, e.g.,][hereafter \citeta{fortney:2007}]{fortney:2007}. After this point, planets and brown dwarfs compress with the addition of mass, such that more massive objects are smaller. Finally, when the pressures and temperatures in the core are sufficiently high, hydrogen burning can begin (marking the start of the stellar regime) and radii again increase with mass. The result is that objects with masses between $\mysim 1$\ and $15$\,\mjup\ are larger than \thisstarb, and those with masses between $\mysim 40$\ and $70$\,\mjup\ are smaller. We must exclude these objects from our posterior (along with smaller planets and larger stars). To do so, we perform rejection sampling using the theoretical mass-radius relationships of \citeta{fortney:2007} for planetary mass objects and MESA models otherwise \citep[those generated for the work in][which extend from $\mysim 2$\ to $100$\,\mjup]{nelson:2018}.

When implementing the \citeta{fortney:2007} models, we assume a core mass of $10$\,\mearth. At the age and insolation of typical companions allowed by the \texttt{EXOFASTv2} model, the predicted radius for a Jupiter-mass object ranges from roughly $1.04$\,\rjup\ (for a core-free planet) to $0.94$\,\rjup\ (for a $50$-\mearth\ core). We therefore adopt a $5\%$\ uncertainty in the predicted radii. We note that for the lowest-mass companions, the radius dependence on the core mass is stronger, and importantly, the opposite is true for more massive planets. If anything, we are allowing relatively too many massive companions to survive the mass-radius selection. 

\begin{figure}[!t]
    \centering
    \includegraphics[width=\linewidth]{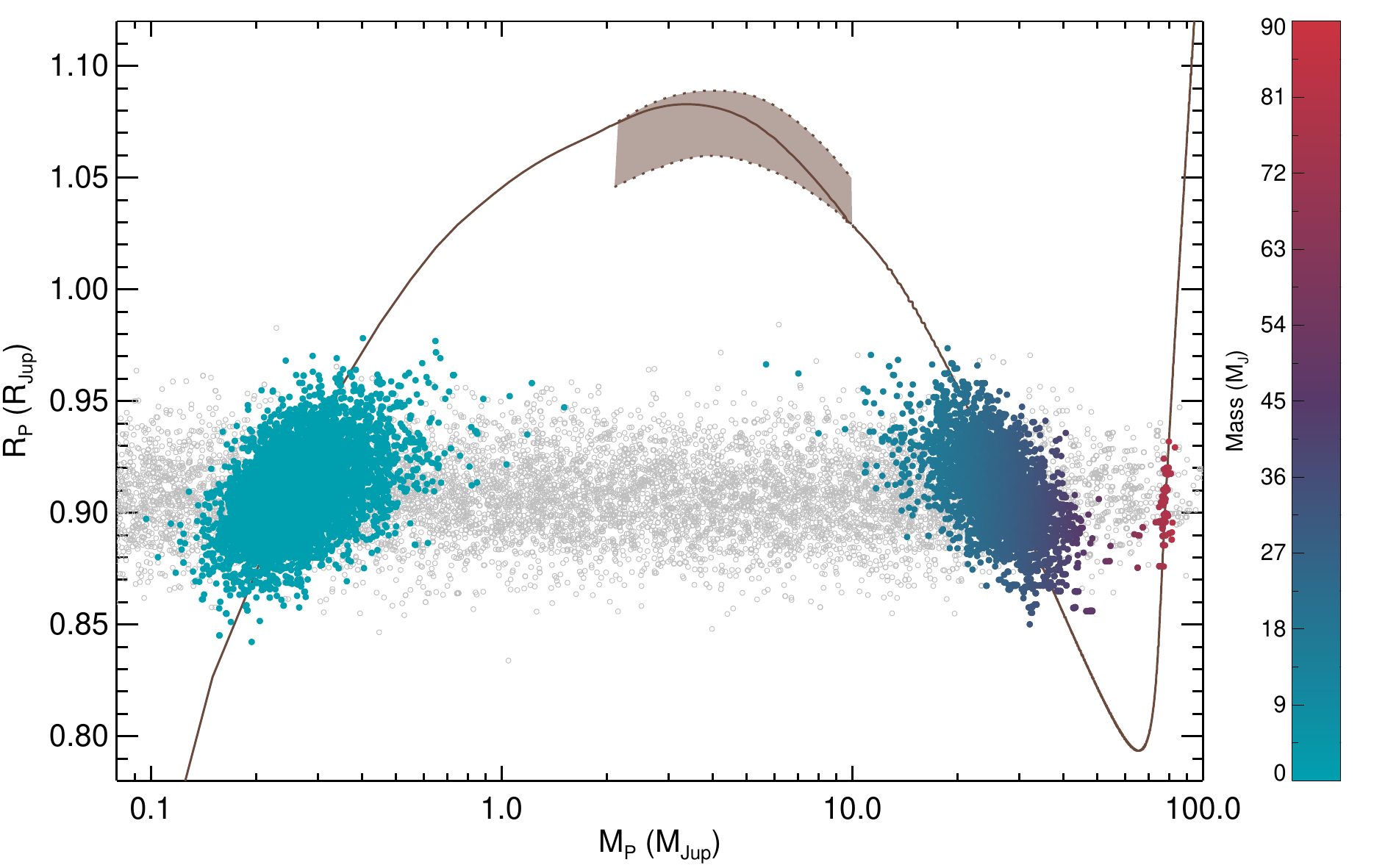}
    \caption{The mass-radius posterior. As described in the text, we reject draws from the \texttt{EXOFASTv2} posterior if they lie far away from mass-radius relations and based on the probability of observing a single transit during the \kep\ mission. The \texttt{EXOFASTv2} posterior is plotted in gray, while the accepted draws from the posterior are colored according to their masses. Brown lines indicate the mass-radius relationship from \citet{fortney:2007} and the MESA models described in \citet{nelson:2018}. The region where the models overlap ($2 < \mpl < 10$\,\mjup) is shaded light brown. The continuous solid brown line indicates our adopted blended model, with each individual model shown as dotted brown lines.}
    \label{fig:mr}
\end{figure}

We also adopt uncertainties of $5\%$\ on predicted radii from the MESA models. This uncertainty encompasses size differences caused by variations in composition and age far beyond our fitted uncertainties for the system, and is therefore a conservative estimate that acknowledges theoretical models of brown dwarf masses and radii may not be free of systematic effects. 

This potential systematic difference is illustrated by the  \citeta{fortney:2007} and MESA models, which overlap between $2 < \mpl < 10$\,\mjup. In this range, MESA predicts smaller radii for a given mass by about $2.5\%$. While we could adjust parameters (such as \citeta{fortney:2007} core mass) to bring the models into agreement, we prefer to use the most realistic \citeta{fortney:2007} models for the planetary companions and accept the inconsistency between models at the few percent level, which is already incorporated in our $5\%$\ uncertainties. To remove discontinuities, we blend the two models in the overlap range by selecting the model used for each sample from the posterior according to a linear distance-weighted probability: near $2$\,\mjup, the \citeta{fortney:2007} models are nearly always selected, and near $10$\,\mjup, the MESA models are nearly always selected. This results in a slightly inflated effective uncertainty on the model radius in this mass range, but ultimately does not affect our results because neither set of models predicts radii in this mass range consistent with the observed radius. A visual representation of the blending of these two models is shown in \rffigl{mr}.

\begin{figure}[!t]
    \centering
    \includegraphics[width=\linewidth]{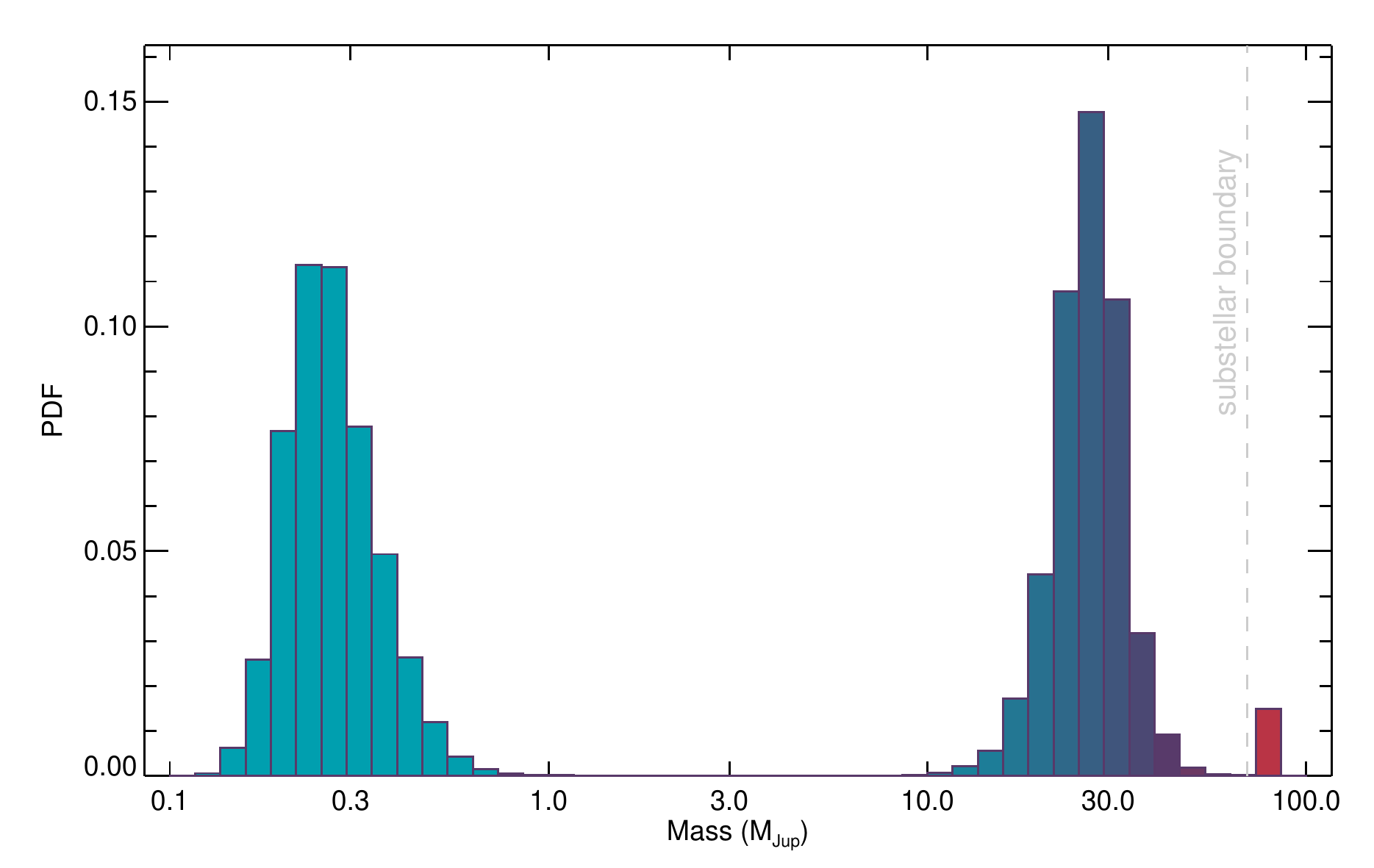}
    \caption{The companion mass posterior, $98.5\%$\ of which is substellar ($\mb \lesssim 70$\,\mjup). The distribution is bimodal, with about $47.4\%$\ of the posterior indicating the companion is a brown dwarf ($\mb = 25.8^{+6.9}_{-4.8}$\,\mjup), and $51.1\%$\ of the posterior indicating a planet ($\mb = 0.238^{+0.084}_{-0.057}$\,\mjup). The true probability of each solution also depends on the relative occurrence rates (and eccentricity distributions) of long period planets and brown dwarfs.}
    \label{fig:mass}
\end{figure}

To determine which draws from the posterior to reject, we assume the uncertainties describe a Gaussian distribution of radii at a given mass, which allows us to calculate analytically the probability of observing a radius as far from the predicted value as each posterior sample. A Gaussian distribution $g(x)\,=\,\frac{1}{\sigma\sqrt{2\pi}}e^{-\frac{1}{2}(\frac{x-\mu}{\sigma})^2}$\ has a cumulative distribution function of the form
\begin{equation}
    CDF = \frac{1}{2} \Bigg(1 + \erf{\Big(\frac{x - \mu}{\sigma\sqrt{2}}\Big)} \Bigg),
\end{equation}
which we apply to calculate the probability of observing each radius in the posterior, $R_i$. That is, given a mass in the posterior, $M_i$, we use the mass-radius relations described above to calculate the predicted radius ($R_{\rm pred}$) and its uncertainty ($\sigma_R$). 
% The CDF of the radii one would expect to observe for that mass $M_i$ is then
% \begin{equation}
    % CDF(x) = \frac{1}{2} \Bigg(1 + \erf{\Big(\frac{x - R_{\rm pred}}{\sigma_R\sqrt{2}}\Big)} \Bigg),
% \end{equation}
% and 
The probability of observing a radius at least as far from $R_{\rm pred}$\ in either direction as $R_i$\ would be $2 \times \big(1 - CDF(R_i)\big)$. We therefore reject each draw from the posterior with the complement of that probability:
\begin{equation}
    {\Pr}_{rej}(R_i) = \erf{\Bigg(\frac{\lvert{R_i - R_{\rm pred}}\rvert}{\sigma_R\sqrt{2}}\Bigg)}.
\end{equation}
The effect of these mass-radius priors can be seen clearly in \rffigl{mr}, and the resulting mass posterior is shown in \rffigl{mass}.

\subsection{Astrometric Constraints from Gaia}
\label{sec:gaia}

If the orbital motion of the host star in the plane of the sky is large, astrometric observations may detect the presence of a companion. Gaia provides the most precise astrometry of any survey to date, but DR2 does not report astrometric orbits. It does, however, report precise 5-parameter solutions (distances, positions, and proper motions). We consider two ways in which a stellar companion might appear in the Gaia data, even in the absence of an astrometric orbital solution.

\subsubsection{Multi-epoch Proper Motions}

If the time span of the astrometric observations is short compared to the orbital period, the derived proper motion may be biased by the orbital motion. If two astrometric measurements are obtained at different epochs, they may therefore show inconsistency if the companion is massive enough to bias the instantaneous proper motions. We can check for differences between the proper motions derived from Gaia data alone (DR2) and  from the combination of PPMXL \citep{roeser:2010} and first-epoch Gaia DR1 positions. DR2 proper motions are derived from observations obtained over the course of about 610 days, while the baseline between PPMXL and DR1 is much longer. Proper motions are derived from these two catalogs and reported in the "Hot Stuff for One Year" (HSOY) catalog \citep{altmann:2017}. Gaia DR2 reports proper motions for \thisstar\ of $\mu_{\rm RA} = -1.400 \pm 0.023$\ and $\mu_{\rm DEC} = -5.319 \pm 0.025\,{\rm mas\,yr}^{-1}$, while HSOY reports $\mu_{\rm RA} = -0.91 \pm 2.10$\ and $\mu_{\rm DEC} = -2.95 \pm 2.10\,{\rm mas\,yr}^{-1}$. These values show disagreement at only the $1\sigma$ level, and it also turns out that the uncertainties on the HSOY proper motions are not precise enough to identify a discrepancy between the multi-epoch proper motions of \thisstar, as the orbital amplitude of the astrometric motion during the time span of DR2 observations, even for the allowed stellar companions is only on the order of $0.1$\,mas yr$^{-1}$. We therefore cannot rule out any massive companions via multi-epoch proper motions.

\begin{figure*}[!t]
    \centering
    \hfill
    \includegraphics[width=0.48\linewidth]{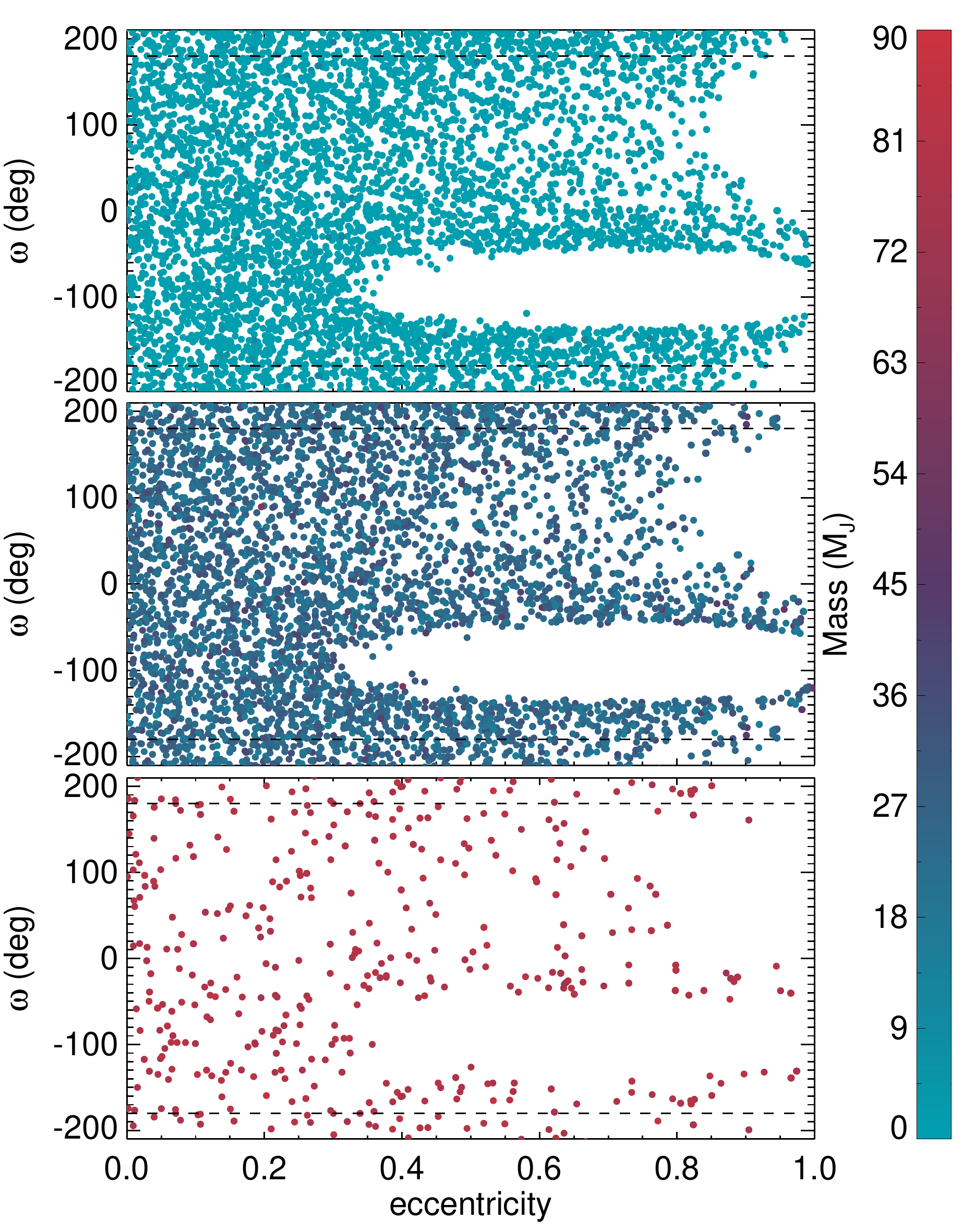}
    \hfill
    \includegraphics[width=0.48\linewidth]{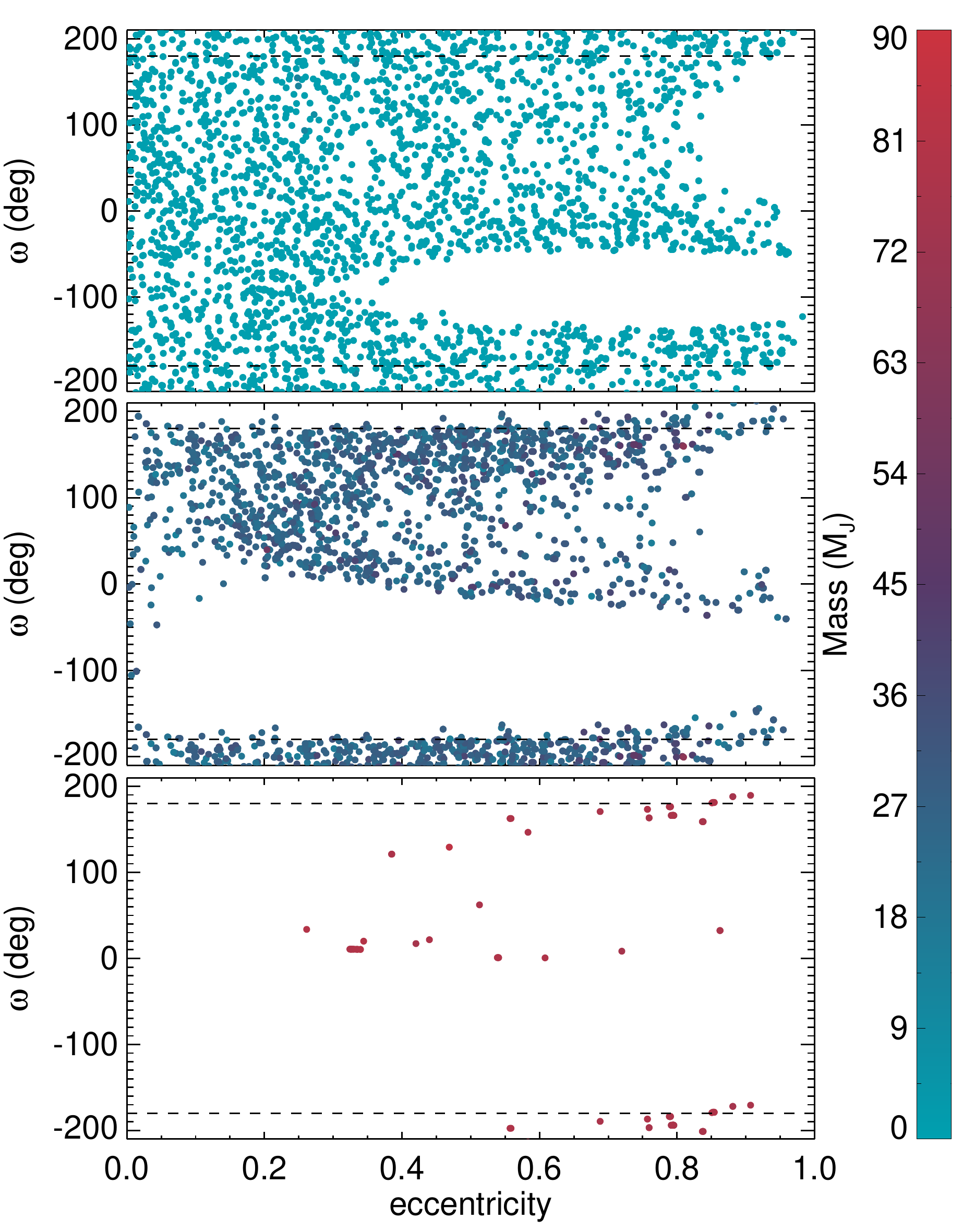}
    \hfill
    \caption{The e-$\omega$ posterior without inclusion of the TRES RVs (left) and with the inclusion of the TRES RVs (right). Points are colored by their mass and separated into planets ($\mb<13\,\mjup$, top), brown dwarfs ($13 < \mb < 70\,\mjup$, middle), and stars ($\mb > 70\,\mjup$, bottom). The void at $e \gtrsim 0.4$\ centered on $\omega=-90\dg$ (corresponding to transit at apastron), can be understood as solutions with transit durations longer than observed for all allowed periods, while the smaller high-eccentricity void centered on $\omega=90\dg$\ corresponds to solutions with transit durations shorter than observed. The RVs rule out the most massive companions, especially for circular orbits and eccentric orbits with the periapse of the planetary orbit pointed away from us.}
    \label{fig:ew}
\end{figure*}

\subsubsection{Astrometric Excess Noise}

In addition to the five-parameter astrometric solution provided in Gaia DR2, an estimate is given of the astrometric excess noise (\aen), which quantifies the additional error term required for a well-fit five-parameter solution. Large \aen\ could result from the presence of a companion causing astrometric motion not well-fit by proper motion and the parallactic ellipse. While the \aen\ for \thisstar\ is formally 0, the presence of a stellar-mass companion cannot be excluded without calculating its expected influence on the astrometry. Even for some massive companions---those for which there was very little orbital {\it acceleration} projected onto the plane of the sky during Gaia observations---the orbital motion could be absorbed into the linear proper motion component of the fit. 
To estimate which solutions are incompatible with $\aen = 0$, for each draw from the posterior, we simulate the DR2 observations under the simplifying assumption that there is no parallactic contribution to the position over time (i.e., the parallax can be fit perfectly). We randomly distribute the $236$\ astrometric measurements over the $610$-day span from BJD 2456892 to 2457532 (Aug 2014 through May 2016), assign per-point uncertainties such that a linear fit for the proper motions yields the uncertainties observed ($\mysim 0.024\,{\rm mas\,yr}^{-1}$), and simulate the measured positions at each time according to the measured proper motions and Gaussian noise. We then calculate the orbital positions of the host star for each of our simulated DR2 times of observation and add them to the simulated positions. Finally, we re-fit the linear proper motions of this new data set. If the orbital motion is significant enough, it will be reflected in the goodness-of-fit. We use the $\Delta\chi^2$\ between the two fits (with and without orbital motion) to reject draws according to the CDF of the $\chi^2$\ distribution: 
\begin{equation}
    {\Pr}_{rej}(\Delta\chi^2) \mysim 1 - e^{-\nicefrac{\Delta\chi^2}{2}}
\end{equation}
\noindent This constraint also turns out to have a minimal effect for \thisstar. Only $0.5\%$\ of draws were rejected by this criterion, most of which would have been rejected anyway by either the mass-radius constraints or the single-transit period prior.

While neither constraint based on DR2 data had an impact for \thisstar, we note that for similar systems discovered orbiting nearby stars, these tests will be more powerful, as astrometric signals will be greater by a factor equal to the distance ratio. This may be important for long-period single-transit systems observed by \tess.

\subsection{The eccentric posterior}
\label{sec:ew}

High quality light curves can be used to place constraints on the orbital eccentricities of transiting planets \citep[e.g.,][]{ford:2008,dawson:2012}. However, this technique relies upon the knowledge of the orbital period (and the stellar properties) to identify deviations in transit duration compared to a circular orbit of the same period and impact parameter. For single transits---i.e., in the absence of orbital period information--- the problem is degenerate. In \rffigl{ew}, we show the joint posterior of the eccentricity and argument of periastron. Without any RV follow-up, very little can be said about the orbital eccentricity. There are two main features in the $e$-$\omega$ distribution: (i) a large void at high eccentricities for transits occurring at apastron ($\omega \mysim -90$\ degrees), corresponding to transits longer than observed, even for the shortest allowed periods; and (ii) a smaller void at high eccentricities for transits occurring near periastron ($\omega \mysim 90$\ degrees), corresponding to transit durations shorter than observed, even for the longest likely periods. Among companions consistent with the observed radius, all masses are allowed.

The inclusion of a handful of RVs across three seasons all but eliminates stellar mass companions, which, apart from a few finely tuned orbits at high eccentricity and long period, should have shown RV variation beyond that which was observed. For brown dwarfs, the RVs have two effects. They extend the void for apastron transits to zero eccentricity, and they rule out many low-eccentricity orbits. This can be understood as a result of the timing of the observations. With RV monitoring beginning $\mysim 5$\ years after the transit and an orbital period roughly twice as long (see \rffigl{period}), a transit at apastron would imply that the RVs were obtained near periastron, when the RVs would be expected to show a large variation if the companion were a brown dwarf. Circular orbits of brown dwarf companions with periods of $\mysim 10$\ years would similarly show large RV variation at the time of our observations. This effect can also be visualized in \rffigl{rv}, which shows RV curves drawn from the posterior. The steepest RV variation (i.e., the periastron passage) of the more massive companions is constrained to be near transit. For planetary mass companions, none of the orbits are expected to be detected in our TRES data, so all orbits allowed by the transit alone remain viable in the final fit.

In the literature, a circular solution is sometimes adopted for single transits, either for simplicity or in order to estimate an approximate period, but as we have shown, radial velocity follow-up---and its timing relative to the transit---can result in a correlation between mass and eccentricity. Examining only circular orbits can therefore be misleading. For example, a circular fit to the \thisstar\ data would lead us to conclude that the companion is very likely to be planetary (with $82\%$\ of the posterior falling near the planetary mass peak) and that the period is well constrained to be 8.2 years regardless of the mass. However, the eccentric fit reveals that neither mass is preferred, the period is likely to be longer, and if the companion is a brown dwarf, a circular orbit is somewhat disfavored.

\subsection{The adopted solution for \thisstar b}

\begin{figure}[!t]
    \centering
    \includegraphics[width=\linewidth]{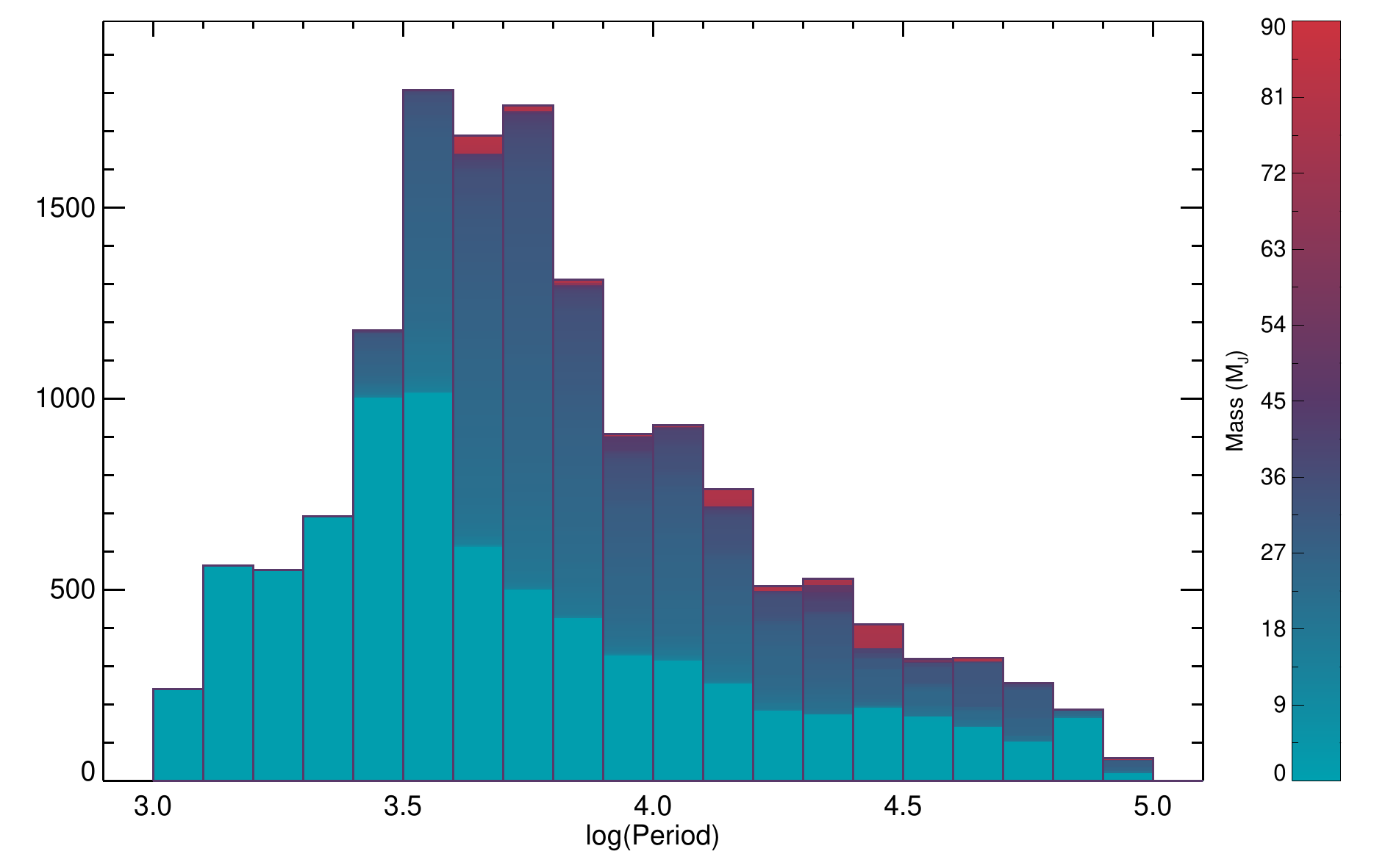}
    \caption{The marginalized posterior of the logarithm of the orbital period, with companions of different masses stacked and colored by their mass. The planetary mass companions (light blue) have an orbital period peaking near $3000$\ days, while the brown dwarf companions (dark blue to purple) peak near $5000$\ days. Both mass regimes include long-period (and high eccentricity) tails. See \rftabl{results} for numerical results.}
    \label{fig:period}
\end{figure}

RV monitoring has not yet ruled out brown dwarf companions, so our posteriors are bimodal (\rffigs{mr}{mass}), with neither solution being strongly preferred by the data. We therefore present two solutions, a brown dwarf companion (encompassing $47.4\%$\ of the probability density) and a planetary companion ($51.1\%$ of the probability density). 
The results of a circular fit, which are provided in \rfsecl{ew} and imply a likely planetary companion, can be used to compare to other single transit systems for which only circular fits are given, but we advise against adopting these parameters. We summarize the results of our eccentric fits in \rftabl{results}. Briefly, we note here the key characteristics of the adopted bimodal solutions. If \thisstarb\ is a planetary companion, it likely has a mass of $0.24$\,\mjup, a period (in $\log{\rm days}$) of $\log{P}=3.49^{+0.42}_{-0.38}$\ ($8.5^{+13.8}_{-5.0}$\ years), and a largely unconstrained eccentricity. If it is a brown dwarf, its likely mass is $26$\,\mjup, the period is longer ($\log{P}=3.70^{+0.38}_{-0.23}$, or $14.7^{+20.6}_{-5.8}$\ years), and the eccentricity is non-zero, with the latter two properties differing from the planetary companions' orbits because RV monitoring has ruled out many short-period and circular brown dwarf solutions. \rffigs{period}{ehist} show the derived period and eccentricity distribution as a function of companion mass.

\begin{deluxetable*}{lccc}
\tabletypesize{\footnotesize}
\tablecaption{Properties of the \thisstar\ system \label{tab:results}}
\tablehead{\colhead{~~~Parameter} & \colhead{Description} & \multicolumn{2}{c}{Values}}
\startdata
\smallskip\\[-2.4ex]
\multicolumn{2}{l}{Stellar Parameters:} & & \\
~~~~$M_*$\dotfill &Mass (\msun)\dotfill &\multicolumn{2}{c}{$1.388^{+0.070}_{-0.088}$}\\
~~~~$R_*$\dotfill &Radius (\rsun)\dotfill &\multicolumn{2}{c}{$1.836\pm0.033$}\\
~~~~$L_*$\dotfill &Luminosity (\lsun)\dotfill &\multicolumn{2}{c}{$4.26^{+0.19}_{-0.18}$}\\
~~~~$\rho_*$\dotfill &Density (cgs)\dotfill &\multicolumn{2}{c}{$0.315^{+0.024}_{-0.027}$}\\
~~~~$\log{g}$\dotfill &Surface gravity (cgs)\dotfill &\multicolumn{2}{c}{$4.052^{+0.026}_{-0.034}$}\\
~~~~$T_{\rm eff}$\dotfill &Effective Temperature (K)\dotfill &\multicolumn{2}{c}{$6119^{+48}_{-49}$}\\
~~~~$[{\rm Fe/H}]$\dotfill &Metallicity (dex)\dotfill &\multicolumn{2}{c}{$0.188^{+0.079}_{-0.077}$}\\
% ~~~~$[{\rm Fe/H}]_{0}$\dotfill &Initial Metallicity \dotfill &$0.244^{+0.083}_{-0.080}$\\
~~~~$Age$\dotfill &Age (Gyr)\dotfill &\multicolumn{2}{c}{$3.03^{+1.1}_{-0.74}$}\\
~~~~$EEP$\dotfill &Equal Evolutionary Point$^{1}$\dotfill &\multicolumn{2}{c}{$393^{+24}_{-22}$}\\
~~~~$A_V$\dotfill &V-band extinction (mag)\dotfill &\multicolumn{2}{c}{$0.122\pm0.040$}\\
~~~~$\sigma_{SED}$\dotfill &SED photometry error scaling\dotfill &\multicolumn{2}{c}{$1.41^{+0.38}_{-0.27}$}\\
~~~~$\varpi$\dotfill &Parallax (mas)\dotfill &\multicolumn{2}{c}{$0.923\pm0.015$}\\
~~~~$d$\dotfill &Distance (pc)\dotfill &\multicolumn{2}{c}{$1082\pm17$}\\
\smallskip\\[-2.4ex]
\multicolumn{2}{l}{Companion Parameters:} & Planets & Brown Dwarfs \\
~~~~$P$\dotfill & Period (days)\dotfill & $2566^{+4155}_{-1682}$ & $4913^{+6562}_{-2758}$ \\
~~~~$\log{{P}}$\dotfill & Log of Period\dotfill & $3.49^{+0.42}_{-0.38}$ & $3.73^{+0.38}_{-0.22}$ \\
~~~~$R_P$\dotfill & Radius (\rjup)\dotfill & $0.909^{+0.017}_{-0.019}$ & $0.911^{+0.016}_{-0.022}$ \\
~~~~$T_C$\dotfill & Time of conjunction (\bjdtdb$-2454833$)\dotfill & $1234.2868^{+0.0085}_{-0.0079}$ & $1234.2823^{+0.0056}_{-0.0069}$ \\
~~~~$a$\dotfill & Semi-major axis (AU)\dotfill & $4.2^{+3.7}_{-2.4}$ & $6.5^{+4.8}_{-2.4}$ \\
~~~~$i$\dotfill & Inclination (Degrees)\dotfill & $89.9920^{+0.0079}_{-0.0170}$ & $89.9881^{+0.0118}_{-0.0080}$ \\
~~~~$ecos{{\omega_*}}$\dotfill & \dotfill & $0.24^{+0.38}_{-0.22}$ & $0.31^{+0.34}_{-0.16}$ \\
~~~~$esin{{\omega_*}}$\dotfill & \dotfill & $-0.000^{+0.015}_{-0.012}$ & $0.0036^{+0.0190}_{-0.0061}$ \\
~~~~$e$\dotfill & Eccentricity\dotfill & $0.22^{+0.37}_{-0.22}$ & $0.30^{+0.35}_{-0.15}$ \\
~~~~$\omega_*$\dotfill & Argument of Periastron (Degrees)\dotfill & $-157^{+218}_{-21}$ & $155^{+24}_{-105}$ \\
~~~~$T_{{eq}}$\dotfill & Equilibrium temperature (K)$^{2}$\dotfill & $184^{+46}_{-62}$ & $152^{+25}_{-42}$ \\
~~~~$K$\dotfill & RV semi-amplitude (m/s)\dotfill & $2.74^{+1.05}_{-0.74}$ & $250^{+77}_{-74}$ \\
% ~~~~$\log{{K}}$\dotfill & Log of RV semi-amplitude\dotfill & $0.46^{+0.15}_{-0.12}$ & $2.42^{+0.11}_{-0.13}$ \\
~~~~$M_P$\dotfill & Mass (\mjup)$^{3}$\dotfill & $0.238^{+0.084}_{-0.057}$ & $25.8^{+6.9}_{-4.8}$ \\
~~~~$M_P/M_*$\dotfill & Mass ratio\dotfill & $0.000163^{+0.000061}_{-0.000040}$ & $0.0183^{+0.0044}_{-0.0038}$ \\
~~~~$R_P/R_*$\dotfill & Radius of planet in stellar radii\dotfill & $0.005214^{+0.000039}_{-0.000033}$ & $0.005208^{+0.000047}_{-0.000031}$ \\
~~~~$a/R_*$\dotfill & Semi-major axis in stellar radii\dotfill & $498^{+429}_{-278}$ & $751^{+576}_{-277}$ \\
~~~~$b$\dotfill & Transit Impact parameter\dotfill & $0.00^{+0.22}_{-0.00}$ & $0.085^{+0.136}_{-0.085}$ \\
~~~~Depth\dotfill &Flux decrement at mid transit$^{4}$ \dotfill &$0.003052^{+0.000025}_{-0.000026}$ &$0.003052^{+0.000025}_{-0.000026}$\\
~~~~$\tau$\dotfill &Ingress/egress transit duration (days)\dotfill &$0.0915^{+0.0068}_{-0.0021}$ &$0.0915^{+0.0068}_{-0.0021}$\\
~~~~$T_{14}$\dotfill &Total transit duration (days)\dotfill &$1.8559^{+0.0079}_{-0.0070}$ &$1.8559^{+0.0079}_{-0.0070}$\\
~~~~$\rho_P$\dotfill & Density (cgs)\dotfill & $0.396^{+0.126}_{-0.087}$ & $41.8^{+13.7}_{-9.8}$ \\
~~~~$logg_P$\dotfill & Surface gravity\dotfill & $2.87^{+0.12}_{-0.10}$ & $4.891^{+0.126}_{-0.085}$ \\
% ~~~~$\fave$\dotfill & Incident flux (\fluxcgs)\dotfill & $-3.57^{+0.46}_{-0.64}$ & $-3.94^{+0.34}_{-0.61}$ \\
~~~~$\fave/\favee$\dotfill & Incident flux relative to Earth\dotfill & $0.19^{+0.37}_{-0.15}$ & $0.085^{+0.101}_{-0.064}$ \\
~~~~$P_{{T,G}}$\dotfill & A priori transit prob\dotfill & $0.00192^{+0.00061}_{-0.00060}$ & $0.00165^{+0.00034}_{-0.00030}$ \\
\smallskip\\[-2.4ex]
\multicolumn{2}{l}{Transit Parameters:}& \multicolumn{2}{c}{Kepler}\\
~~~~$u_{1}$\dotfill &linear limb-darkening coeff\dotfill &\multicolumn{2}{c}{$0.348\pm0.032$}\\
~~~~$u_{2}$\dotfill &quadratic limb-darkening coeff\dotfill &\multicolumn{2}{c}{$0.272\pm0.045$}\\
~~~~$\sigma^{2}$\dotfill &Added Variance\dotfill &\multicolumn{2}{c}{$1.13^{+0.22}_{-0.20} \times 10^{-8}$}\\
~~~~$F_0$\dotfill &Baseline flux\dotfill &\multicolumn{2}{c}{$1.000003\pm0.000011$}\\
\smallskip\\[-2.4ex]
\multicolumn{2}{l}{Instrument Parameters:}&\multicolumn{2}{c}{TRES}\\
~~~~$\gamma_{\rm rel}$\dotfill &Relative RV Offset (m/s)\dotfill &\multicolumn{2}{c}{$36^{+37}_{-30}$}\\
~~~~$\sigma_J$\dotfill &RV Jitter (m/s)\dotfill &\multicolumn{2}{c}{$38\pm38$}\\[1ex]
% ~~~~$\sigma_J^2$\dotfill &RV Jitter Variance \dotfill &\multicolumn{2}{c}{$1500^{+4400}_{-1900}$}\\[1ex]
\enddata
% \tablenotetext{}{\scriptsize See Table 3 in \citet{eastman:2019} for a detailed description of all parameters.\\[-4ex]}
% \tablenotetext{1}{\scriptsize Corresponds to static points in a star's evolutionary history. See \S2 in \citet{dotter:2016}.\\[-4ex]}
% \tablenotetext{2}{\scriptsize Assumes no albedo and perfect heat redistribution.\\[-4ex]}
% \tablenotetext{3}{\scriptsize The mass posteriors are shaped by the observed RV variation and the importance sampling of mass-radius relationships as described in \rfsecl{mr}.\\[-4ex]}
% \tablenotetext{4}{\scriptsize Observed depth in the \kep\ bandpass. This is greater than $(R_P/R_*)^2$\ because of limb darkening.}
\scriptsize See Table 3 in \citet{eastman:2019} for a detailed description of all parameters.\\
$^1$ Corresponds to static points in a star's evolutionary history. See \S2 in \citet{dotter:2016}.\\
$^2$ Assumes no albedo and perfect heat redistribution.\\
$^3$ The mass posteriors are shaped by the observed RV variation and the importance sampling of mass-radius relationships as described in \rfsecl{mr}.\\
$^4$ Observed depth in the \kep\ bandpass. This is greater than $(R_P/R_*)^2$\ because of limb darkening.

\end{deluxetable*}

\section{Discussion}
\label{sec:discussion}

\begin{figure}[!t]
    \centering
    \includegraphics[width=\linewidth]{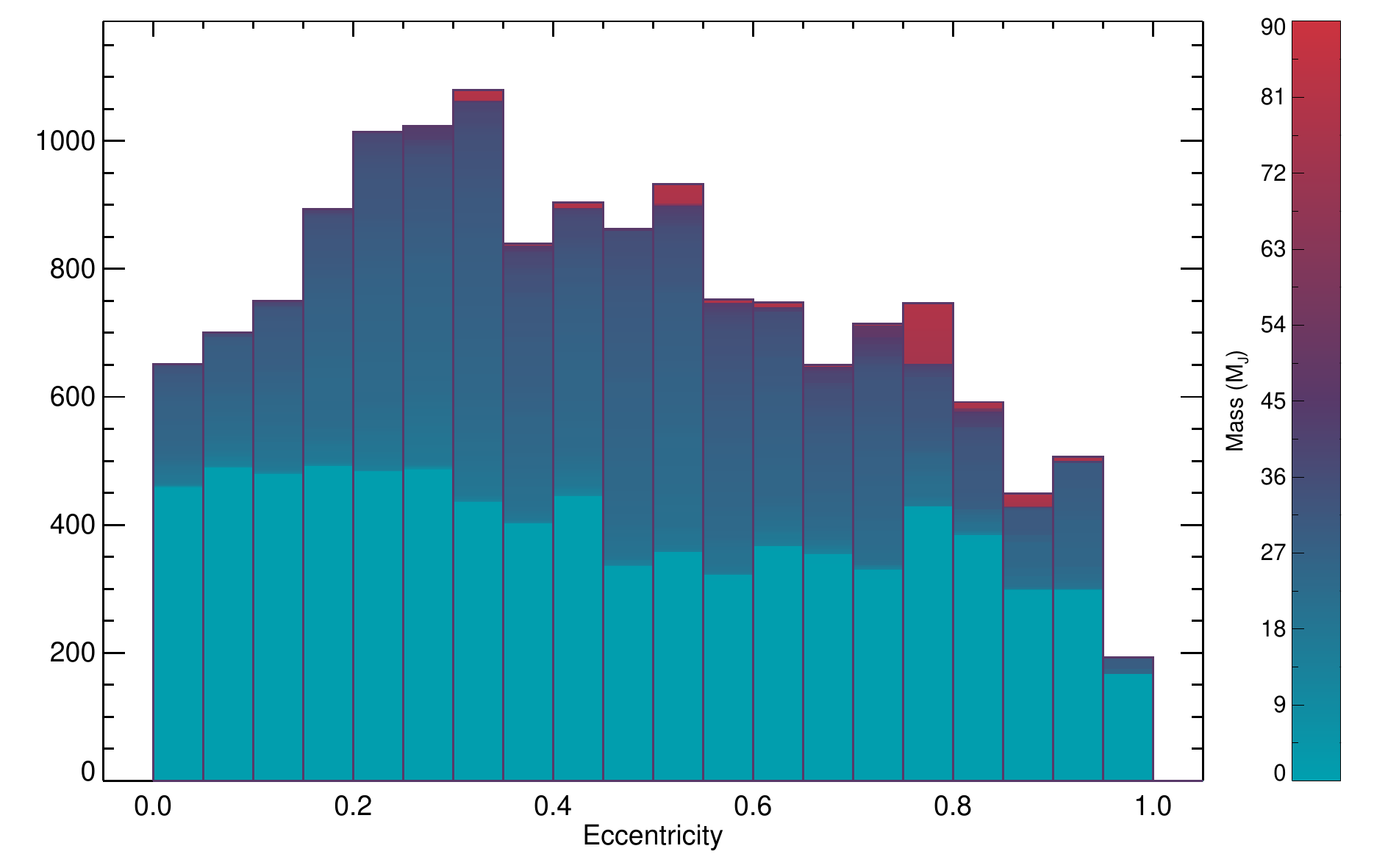}
    \caption{The marginalized posterior of the orbital eccentricity, with companions of different masses stacked and colored by their mass. As seen in \rffigl{ew}, the eccentricity of planetary mass companions (light blue) is largely unconstrained, while the brown dwarf companions (dark blue to purple) are preferentially eccentric. See \rftabl{results} for numerical results.}
    \label{fig:ehist}
\end{figure}

\begin{figure}[!t]
    \centering
    \includegraphics[width=\linewidth]{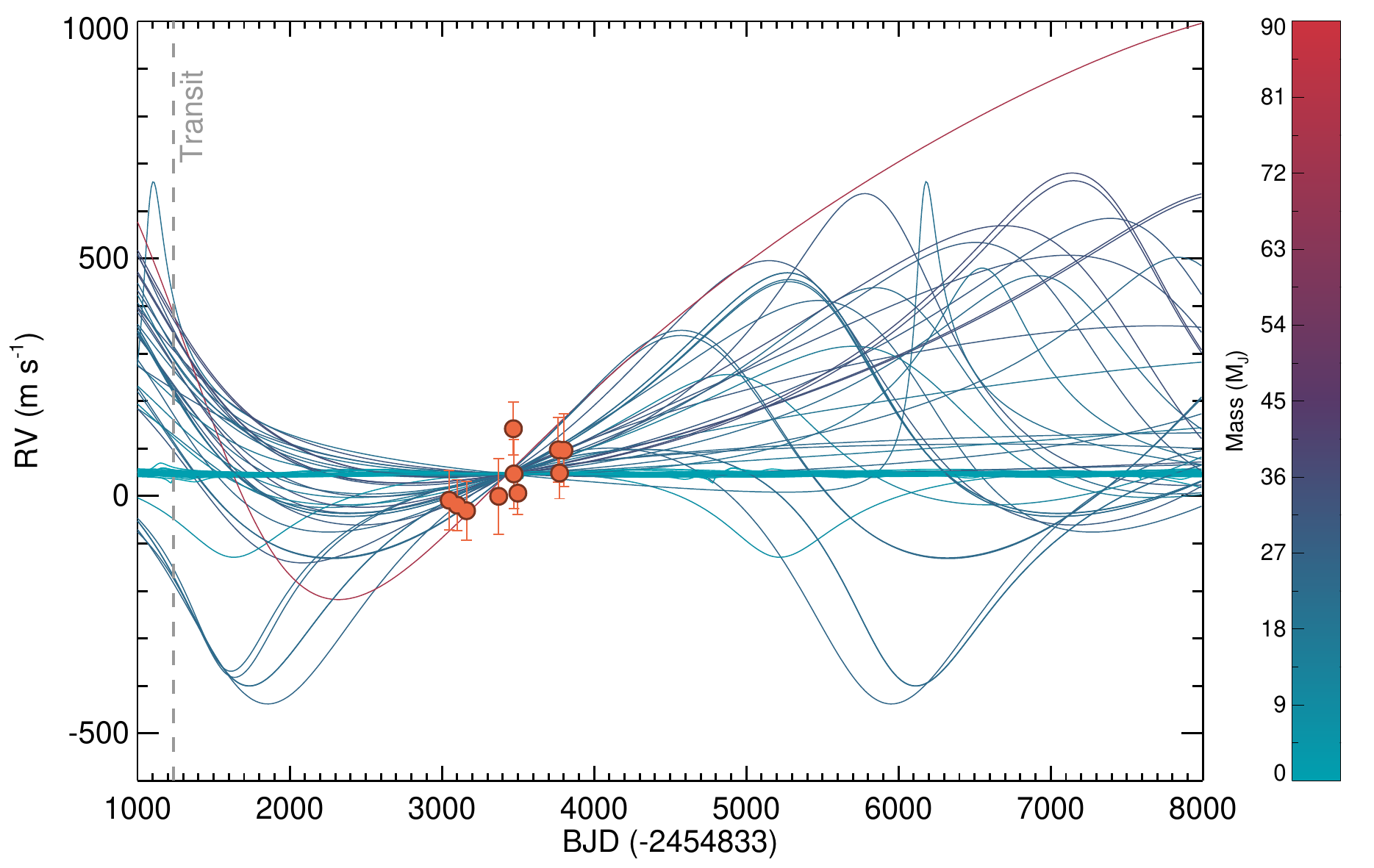}
    \caption{TRES radial velocities of \thisstar, shown in orange. The time of transit is indicated by the vertical dashed gray line. We plot 100 RV curves drawn at random from the posterior, color-coded by the companion mass. The brown dwarf solutions tend to be eccentric and long-period, but most importantly, have semi-amplitudes of hundreds of \ms, which would be detectable with continued RV monitoring.}
    \label{fig:rv}
\end{figure}

The properties of \thisstarb, while qualitatively constrained to be long-period and substellar, remain uncertain. Continued RV monitoring does have the potential to measure the mass and period if the companion is a brown dwarf. This can be seen in \rffigl{rv}, in which the RV curves of the planetary solutions remain flat to within the measurement errors, while the brown dwarf solutions diverge over the next several years, with amplitudes of hundreds of \ms. Additional RVs should collapse the posteriors to a unimodal solution. In the event that \thisstarb\ is massive enough to induce a detectable RV variation, the measurement of its mass, radius, and period would result in a valuable brown dwarf benchmark; because the host star has left the main sequence, its age is also well constrained ($3.03^{+1.2}_{-0.68}$\,Gyr). There are no similar brown dwarf benchmarks. 

A growing number of brown dwarfs have been detected via transit \citep[see, e.g.][]{carmichael:2020,carmichael:2021}, yielding model-independent masses and radii, but these systems have thus far been limited to short periods, indicating a difference in formation and/or migration histories, as well as elevated irradiation. Long-period and isolated brown dwarfs have been directly imaged, but their physical parameters must be constrained by theoretical models and (often imprecise) age estimates. Many long-period brown dwarfs do have measured astrometric orbits and precise dynamical masses \citep[see, e.g.,][]{dupuy:2017}, but derived radii are still model-dependent. Measurements of the radii from transits would be important for testing and calibrating evolutionary models.

Additional observations of a long-period companion \textit{during} transit---to directly detect its atmospheric via transmission spectroscopy or its orbital alignment via the Rossiter--McLaughlin effect or Doppler tomography---would be extremely valuable, and though it would be difficult, it is possible that a dedicated spectroscopic and photometric campaign could succeed in recovering a second transit. After RV monitoring to narrow the transit window, sparse photometric observation, even over a long time span, would be feasible. The duration of the event is long enough that low data points detected on one night could act as a trigger for intensive egress observations over the next two nights. (It is unrealistic to expect successful transit recovery if \thisstarb\ is the mass of Saturn because the RVs will not help refine the period estimate.) Unfortunately, characterization of \thisstarb\ through other means common for brown dwarfs is unlikely because of the distance to the system. Even if it were a brighter star and a good target for future imaging facilities like the Nancy Grace Roman Space Telescope, inner working angles as small as $0\farcs05$ would only detect objects at separations greater than 50 au at a distance of $1$\,kpc. The astrometric signal of \thisstar\ is likely to be too small to be detected as well. \citet{ranalli:2018} estimate that the maximum distance for detection of a $1$\,\mjup\ companion orbiting at $4$\,au in a 10-year Gaia mission is $\mysim 70$\,pc. This detection limit scales to roughly the mass and distance of \thisstarb\ in the brown dwarf scenario, but does not take into account the likely longer period. The orbit size would therefore be larger, but Gaia would likely observe only a partial orbit and may struggle to detect it. 

Because of its distance, \thisstarb\ is not a substellar, long-period transiting object that is well-suited for detailed characterization, but we may soon find similar systems more amenable to follow-up, and the techniques described herein can be applied to brighter, closer systems, such as those expected from \tess. From their comprehensive single transit search in \kep\ data, \citet{kawahara:2019} estimate an occurrence rate of $0.39$\ planets per FGK star with orbital separations between $2$\ and $20$\ au and radii between $4$\ and $14$\,\rearth. This is not inconsistent with the occurrence rate of long-period giant planets derived from radial velocities \citep{fulton:2021}. While \tess\ stars are only observed for between 27 days and one year during the two-year prime mission, \tess\ observes orders of magnitude more stars than \kep. If we take the \tess\ Candidate Target List (version 8; CTL-8), for example, which contains roughly 10 million stars ranked highly based on metrics that prioritize transit detection \citep{stassun:2018b,stassun:2019}, we can use the CTL Filtergraph portal\footnote{\url{ http://filtergraph.vanderbilt.edu/tess_ctl}} \citep{filtergraph} to select FGK stars ($4700 < \teff < 6500$\,K) brighter than \thisstar. We find $3.6$\ million results. Assuming the average timespan of observation is $\mysim 54$\ days (2 \tess\ sectors), the integrated \tess\ observing time for these bright FGK stars will roughly equal the integrated observing time for all \kep\ stars. We note that this ignores the ongoing \tess\ extended mission, which will more than double the data volume from the prime mission, and provide light curves for many more hot stars and cool dwarfs than \kep. 

Given the yield from \kep\ \citep[67;][]{kawahara:2019}, we therefore expect that a comprehensive search of \tess\ stars for single transits should return dozens of planets in wide orbits around FGK stars brighter than \thisstar, for which follow-up will be possible. If we restrict our selection to the brightest FGK stars ($T<10$), we still expect a couple planets similar to \thisstarb\ orbiting very bright FGK stars from the \tess\ prime mission, with more in the extended mission and still more orbiting other types of stars. These predictions are also broadly consistent with those of \citet{villanueva:2019}, who used a target list of $4$\ million stars with a wide range of effective temperatures to estimate \tess\ single transit detections at all orbital periods. They do not report results for the same parameter space, but among FGK stars, they estimate 18 habitable zone planets even though FGK stars make up only a fraction of their target list (their Figure 2). For a target list similar to ours ($3.6$\ million FGK stars), their long-period planet yield should be a factor of a few larger. 

While bright hosts are important for improving RV precision, it is the combination of brightness and proximity to Earth of the predicted \tess\ long-period single-transit hosts that will enable complementary characterization techniques such as astrometry and direct imaging. The sample described above has a typical distance of $300$--$400$\ pc, while the brightest few candidates will come from a distribution with typical distances of $\mysim 100$\ pc. At $100$\ pc, astrometry and direct imaging can plausibly detect signals arising from these companions, even for planetary masses, leading to a more complete characterization of their physical properties and orbits. These single-transit systems therefore hold great promise to become benchmark objects among companions formed near the snow lines of Sun-like stars.

\section{Conclusion}
\label{sec:conclusion}

We identify a companion to the star \thisstar\ that exhibits a single, 45-hour transit in the \kep\ data, implying an orbital period similar to the gas giants of the Solar System. The size of the companion is consistent with objects the mass of Saturn, brown dwarfs, or low-mass stars, so cannot be characterized by photometry alone. We model the host star, the transit, and three seasons of radial-velocity data to show that the companion is substellar. While the assumption of a circular orbit would imply a likely Saturn-mass object with a period of 8 years, the data can also be fit with an eccentric orbit, in which case the solution is bimodal, nearly equally well fit by Saturn-mass objects with a period of $\mysim 10$ years or brown dwarfs about 26 times the mass of Jupiter with periods twice as long. Additional RV follow-up should distinguish between the two solutions, and may lead to an orbital solution if the companion is massive, providing a benchmark long-period brown dwarf with a measured mass, radius, and age. We expect that \tess\ will discover similar objects orbiting bright, nearby host stars more amenable to follow-up with RVs, astrometry, and direct imaging.

% \vfill
% \vspace{1.1in}

\begin{acknowledgments}

This paper includes data collected by the Kepler mission and obtained from the MAST data archive at the Space Telescope Science Institute (STScI). Funding for the Kepler mission is provided by the NASA Science Mission Directorate. STScI is operated by the Association of Universities for Research in Astronomy, Inc., under NASA contract NAS 5-26555.

This work has also made use of data from the European Space Agency (ESA) mission {\it Gaia} (\url{https://www.cosmos.esa.int/gaia}), processed by the {\it Gaia} Data Processing and Analysis Consortium (DPAC, \url{https://www.cosmos.esa.int/web/gaia/dpac/consortium}). Funding for the DPAC has been provided by national institutions, in particular the institutions participating in the {\it Gaia} Multilateral Agreement.

Some of the data presented herein were obtained at the W. M. Keck Observatory, which is operated as a scientific partnership among the California Institute of Technology, the University of California and the National Aeronautics and Space Administration. The Observatory was made possible by the generous financial support of the W. M. Keck Foundation. The authors wish to recognize and acknowledge the very significant cultural role and reverence that the summit of Maunakea has always had within the indigenous Hawaiian community.  We are most fortunate to have the opportunity to conduct observations from this mountain. 

This research has made use of NASA's Astrophysics Data System Bibliographic Services.

\end{acknowledgments}

\facilities{Kepler, FLWO:1.5m (TRES), Keck:I (HIRES), Gaia}

% \vspace{1in}
% \vfill

\software{\texttt{EXOFASTv2} \citep{Eastman:2017,eastman:2019}}

\bibliographystyle{aasjournals}

\bibliography{kic491}

\end{document}